\documentclass[a4paper]{article}

\newcommand{\dcp}{\delta_{CP}}
\newcommand{\dmsq}{\Delta m^2}
\newcommand{\dmsqc}{\Delta m^2_{32}}

\newcommand{\dmsqa}{\Delta m^2_{21}}
\newcommand{\dmsqb}{\Delta m^2_{31}}
\newcommand{\dmsqaz}{\Delta m^2_{12}}
\newcommand{\dmsqi}{\Delta m^2_{ij}}
\newcommand{\qa}{\theta_{12}}
\newcommand{\qb}{\theta_{23}}
\newcommand{\qc}{\theta_{13}}
\newcommand{\znb}{0\nu\beta\beta}
\newcommand{\etal}{\it et al.}

\usepackage{16lomcon}        
\usepackage{cite}             

\begin{document}


\title{Neutrino Physics in 2020}

\author{Maury Goodman \email{maury.goodman@anl.gov}}
\affiliation{Argonne National Lab, Argonne IL 60439, USA}
\date{}
\maketitle
\begin{abstract}
Many talks at the 16th Lomonosov Conference, dedicated to
Bruno Pontecorvo, detail the remarkable progress in neutrino
physics over the last two decades.  In this paper, I give
an opinionated, and therefore likely inaccurate, review of
the future, with some opinions on how both the physics situation
and future facilities will develop, focusing on the year 2020.
\end{abstract}
\section{Introduction}
There are many other papers in these proceedings describing 
results from past experiments, new projects and ideas for 
furthering our knowledge of the neutrino.
There has been a great blossoming of results in the last 20 years 
which shed light on the properties of the neutrino, mostly based
on the phenomenon of neutrino oscillation, as first described by
Pontecorvo 46 years ago \cite{bib:pontecorvo}.
\section{Notation and Semantic Issues}
It is common to see the standard model particles listed from a Particle
Data Group (PDG) table  made in the last century.  The 
neutrinos are listed as $\nu_e$, $\nu_\mu$ and $\nu_\tau$.  
Cabbibo mixing distinguishes the s and s$'$ states, but the CKM
matrix is mostly diagonal, so we usually use
the same 6 quark
labels to describe the flavor and mass eigenstates.
The quarks never appear singly anyway.
For neutrinos, we now know,
the mass and flavor states are quite different.  Which is the particle?
We usually imagine a particle as being able to 
travel from point to point.  The solution to 
the vacuum Schrodinger equation is a mass eigenstate,
so the ``particles" are $\nu_1$, $\nu_2$
and $\nu_3$.  Since we don't yet know the order, and families are
usually separated by their mass (also the order in which we
found them), the PDG chart now labels them 
as $\nu_{lightest}$, $\nu_{middle}$ and $\nu_{heaviest}$.
The chart can be updated when the hierarchy is known.  
If we happen to be in the inverted hierarchy, the order may seem
strange, and there may be reasons to rename
everything.
But a list of particles with the neutrino flavor states should be
corrected.
The hierarchy can be measured with
several different techniques, so it is the focus of attention
from a large fraction of the $\nu$ community. 
The hierarchy is equivalent to the sign of
$\dmsqc$.  Most physicists define $\dmsqi \equiv m_i^2 - m_j^2$.
It is infrequent though possible to define $\dmsqi \equiv m_j^2 - m_i^2$.
But if we see $\dmsqaz$ when we think
they mean $\dmsqa$, it is more likely a mistake.  
The subscripts on the mixing angles, $\qa$, $\qb$ and $\qc$
are just labels, and the order could have been anything.  But since
the 2-$\nu$ mixing approximation has been so useful, some
people (wrongly) associate $\dmsqaz$ with $\qa$.  Why does any of
this matter?  For the $\znb$ the hierarchy is crucial.
But at this meeting in the talk
on $\znb$ decay, a graph was shown in which $\dmsqc$
labels were exactly backwards.\cite{bib:bara}
%
\par Another question is whether
or not neutrino mass is ``physics beyond the standard model."  While
early versions of the standard model explicitly made the neutrino
masses zero, since they thought that they were, it is trivial to
add Dirac neutrino masses.  It
can be said that this would require a new discrete symmetry
to prevent the appearance of Majorana mass.  That is new.
But the majority of theorists whom I have
asked state that neutrino mass in and of itself is not beyond the
standard model.  They mean it is unrelated to the
issues that technicolor, supersymmetry, extra dimensions, and other
ideas were designed to solve.
\section{Inside and outside the 3 $\nu$ paradigm in 2013}
\begin{table}[hbt] 
\centering 
{\begin{tabular}{|c|c|c|}
\hline 
& value & error \\ \hline \hline
$\sin^2(2\qa)$ & 0.857 & 0.024 \\
$\sin^2(2\qb)$ & $>$0.95 & \\
$\sin^2(2\qc)$ & 0.095 & 0.010 \\
$\dmsqa$ & $7.5 \times 10^{-5}eV^2$ & 0.020 \\
$|\dmsqc|$ & $2.32 \times 10^{-3}eV^2$ & 0.012 \\
\hline 
\end{tabular}} 
\caption{Neutrino Mixing Parameters from 2013 PDGLive.}
\label{tab:par} 
\end{table}
What do we know about neutrinos in 2013?  Through
the energies of a large variety of solar, atmospheric, reactor and
accelerator experimenters, we now know, within the 3-$\nu$
paradigm, three mixing angles, both values of $\dmsq$ and the sign
of one of them.  These are listed
in Table \ref{tab:par}.  Many of the beautiful experiments which helped contribute
to this knowledge are described in these proceedings\cite{bib:16lom}.
Outside the 3-$\nu$ paradigm, these numbers may be approximations or
meaningless.  
The currently unknown aspects of the 3-$\nu$ paradigm can be listed as
follows:
(A) What is the mass hierarchy or sign of $\dmsqc$?
(B) Is $\qb$ maximal and if not, which octant is it in?
(C) What is the value of the CP violation parameter $\dcp$?
(D) What is the overall mass scale?
(E) Is the nature of the neutrino Dirac or Majorana?
(F) We would like to know the parameters in Table \ref{tab:par} more accurately.
\section{The mass hierarchy and CP violation \label{sec:mass}}
There are several ways  
 to measure the hierarchy.  In accelerator experiments, 
matter effects provide a difference between $\nu$ and
$\bar{\nu}$ rates which differ from those due to CP violation.  Reactor
experiments, which measure a different mixture of $\dmsqb$ and $\dmsqc$
than accelerator experiments, could determine which one is larger with a
large detector at 50 km with good energy resolution.  JUNO and RENO50
are planning to do that.  Atmospheric neutrino experiments are sensitive
to matter effects through the angular distributions of
$\nu_\mu$,  $\bar{\nu}_\mu$ and $\nu_e + \bar{\nu}_e$.  The
PINGU collaboration's proposal for an add-on to Ice-Cube would
use that signal, as would the India-Based Neutrino Observatory's magnetized
ICAL detector.  An observation of a supernova could provide the answer if
a spectrum-swap is seen, i.e. a time dependent change in the $\nu_\mu$ and
$\nu_e$ energy distributions due to an MSW-like effect in $\nu \nu$ scattering.
But cosmology fits could be the first to provide the answer, since 
the sum of neutrino masses 
$\Sigma_i (m_i^\nu)$ 
in the normal (inverted) hierarchy is 
$>$ 55 ($>$ 105)meV.
In March 2013 Planck
limits this sum to no more than 230 meV \cite{bib:planck} and the error on this sum could reach
50 meV by 2019 with the measurement of B field polarization
in the cosmic microwave background.  This method could determine
we are in the normal hierarchy, but cannot distinguish the non-hierarchical
normal hierarchy from the inverted hierarchy.  Keep one thing
in mind about the mass hierarchy.  When we measure $\qb$, $\dcp$ or the
mass of the Higgs, there is value in trying to measure it better.  When
we measure the mass hierarchy, there is nothing to measure better.  And
while we often want to measure things in different ways as a consistency
check on our 3-$\nu$ paradigm, I doubt that if additional physics
exists in the neutrino sector, that it will manifest itself as different
answers for the hierarchy.
\par The best way to measure CP violation will be using electron neutrino
appearance in sufficiently large long-baseline accelerator experiments.
NO$\nu$A and T2K will make the next measurements in this decade until
larger experiments are built in the next.
Long distances aren't required, but such future 
experiments will be put at a long
enough baseline to assure resolution of the mass hierarchy.  
The candidate
programs are LBNE in the U.S. (favored by the recent ``Snowmass" meeting
of the HEP community), Hyper-K in Japan, and LBNO in Europe.
From a funding/politics point of view, the latter program seems unlikely.
Each program has been described at this conference 
\cite{bib:lbne,bib:lbno,bib:hyperk}.  These are sufficiently long-term
beam and detector construction projects that it will be beyond 2020
before any exist and have results. 

\section{The Challenge of neutrinoless double beta decay ($\znb$)}
If the neutrino is Majorana, $\znb$ decay happens at
predictable rates depending on masses, mixing angles and matrix elements.
If the neutrino is Dirac, it does not happen.  Matrix element calculations
are difficult, but the differences between calculations 
are at most a factor of four.  Experiments will soon be testing rates
at a level that will soon be sensitive to part of the available parameter
space for the inverted hierarchy, but since sensitivity improves only
with the fourth root of statistics, covering that entire parameter space
will take some time.  Sensitivity to the hierarchical normal hierarchy
is in the distant future.  In that difficult case, we can contrast the
sensitivity of $\znb$, which is sensitive to $\Sigma_i U_{ei}^2 m_i$
to tritium beta decay sensitive to $\Sigma_i U_{ei}^2 m_i^2$.  We now know the
central value of everything but $m_1$, but since $m_2 > m_1$ we can
calculate both sums.  In this interesting case, 
$\znb$ decay is dominated
by $m_2$, and tritium beta decay is dominated by $m_3$.
\section{Predictions and prognosis for 2020}
Where will the neutrino world
be in the year 2020?  I think we will have measured the
hierarchy.  If $\qb$ isn't too close to $\pi/4$, 
we will know its quadrant.  The errors listed
for the parameters in Table~\ref{tab:par} will be smaller.
We'll have more
information on $\dcp$, but unless we are
quite lucky, we won't have satisfactorily 
established CP violation.  
And we won't yet know whether 
the neutrino is Dirac or Majorana.  
\par I suspect we actually will
know something about the overall mass scale, from
cosmology experiments as mentioned in Section~\ref{sec:mass}.
Those searches involve fits of cosmological parameters, including
neutrino mass.  Since those fits are
model dependent, there may be effects outside the currently
favored $\Lambda$CDM model which mimic neutrino mass.  The challenge 
for the particle physics community will be to evaluate the
fits and their assumptions and decide whether to accept them, since they are
based in part on issues beyond our expertise.  But as the
solar neutrino problem taught us, it would be a mistake to
discount answers that come from fields we don't totally
understand.   
\par The remaining program in 2020 will be to measure $\dcp$
to determine whether there is CP violation.
We'll be able
to determine the Dirac/Majorana nature of the neutrino in a short time if
the hierarchy is inverted, but in a much longer time frame 
for the normal
hierarchy.  
Will it be important to improve measurements of the 
parameters in Table I?
The answer depends on the
state of neutrino theory at the time.  If theorists
can find any possible sense in these values, 
then more accurate measurements might be needed.
If the theoretical situation on 
masses and mixing angles hasn't changed, then extraordinary efforts
to measure neutrino parameters better don't seem justified to me.
The exceptions are if $\qc$ is still $\pi$/2 within errors or
$\dcp$ is indistinguishable from 0 or $\pi$. 
\par I've been assuming the validity of the 3-$\nu$
paradigm.  Some ``anomalies" have
stimulated interest in new experiments to search for sterile
neutrinos with an eV scale mass.  This reminds us that
there is always the possibility that these or other experiments will
find something in the neutrino sector beyond our current understanding
of 3-$\nu$ mixing.  That would point us toward
new experiments.  I don't know
what those new experiments would be.
\section{Neutrino miracles}
The conference participants had a tour of Sergeiv Prosad/Zagorsk.  
Several miracles were
described to us there.  One definition of a miracle
could be whenever we have a failure of Murphy's Law \cite{bib:murphy}.
A collection of observations about the
seeming ``intelligent design" of neutrino
properties from about ten years ago is worth recounting \cite{bib:stan}:
{\it 1) The optimum choice for $\dmsqa$?  --                                   
Such as to give resonant transition (MSW effect) 
in the middle of solar energy spectrum;
2) The optimum choice for $\qa$? --  Big enough for oscillations 
to be seen in KamLAND;
3) The optimum choice for $\dmsqc$? --                                    
Such as to give full oscillation in the middle of the range of 
possible distances that atmospheric $\nu$s travel to get to the detector;
4) The optimum choice for $\qb$? -- 
Big enough so that oscillations could be seen easily;
5) The optimum choice for $\qc$? --                                    
Small enough so as not to confuse interpretation of the above;
But the acid test is will $\qc$ be big enough to see CP violation 
and determine the hierarchy?  }
The last condition has been famously met \cite{bib:q13}.
Let me extrapolate to the miracles we might add to this list by
2020.  At the risk of being too greedy, we would like to see our
program move forward, and that will happen best with:
1)$\dcp \sim \pi/2$
  to most quickly determine the hierarchy and
  to get large CP violation;
2) The inverted hierarchy, so we can tell Dirac/Majorana and 
maybe the beta decay endpoint;  
3) Majorana, which seems to be more interesting to theorists, 
and we want our theorists to be happy.
\section*{Acknowledgments}
Thanks to the organizers of the Lomonosov meeting, and in 
particular to Prof. Alexander I.Studenikin
for creating a great series of workshops.

\end{document}